\documentclass[10pt]{article}   

\usepackage{amsthm,amsmath,amssymb,mathrsfs,cite,amscd}

\newcommand{\p}{\partial}

\newcommand{\CC}{\mathcal{C}}

\newcommand{\CO}{\mathcal{O}}

\renewcommand{\geq}{\geqslant}

\newcommand{\C}{\mathbb{C}}
\newcommand{\N}{\mathbb{N}}
\newcommand{\R}{\mathbb{R}}

\newcommand{\Z}{\mathbb{Z}}

\renewcommand{\P}{{\bf P}^1} 

\newcommand{\bs}{\begin{split}}
\newcommand{\be}{\begin{equation}}
\newcommand{\es}{\end{split}}
\newcommand{\ee}{\end{equation}}

\begin{document}

\theoremstyle{plain}
\newtheorem{definition}[equation]{Definition}
\newtheorem{lem}[equation]{Lemma}
\newtheorem{prop}[equation]{Proposition}
\newtheorem{thm}[equation]{Theorem}
\newtheorem{conj}[equation]{Conjecture}
\newtheorem{cor}[equation]{Corollary}

\theoremstyle{remark}
\newtheorem{rem}[equation]{Remark}
\newtheorem{ex}[equation]{Example}



\begin{titlepage} 
  \linespread{1.8}
  \title{\Large \bf Integrality of Open Instantons Numbers}
  \author{Daniel B. Gr\"unberg \\ {\em \small KdV Institute, Plantage
  Muidergracht 24, 1018 TV Amsterdam, The Netherlands} \\ \small grunberg@science.uva.nl} 
  \date{May 2003}
  \maketitle
  \vspace{2cm}
  \begin{abstract} \large
We prove the integrality of the open instanton numbers in two examples
of counting holomorphic disks on local Calabi-Yau threefolds: the
resolved conifold and the degenerate $ \P \times \P $.  Given the
B-model superpotential, we extract by hand all Gromow-Witten
invariants in the expansion of the A-model superpotential.  The proof
of their integrality relies on enticing congruences of binomial
coefficients modulo powers of a prime.  We also derive an expression
for the factorial $(p^k-1)!$ modulo powers of the prime $p$.
We generalise two theorems of elementary number theory, by
Wolstenholme and by Wilson. 
  \end{abstract}
  \thispagestyle{empty}
\end{titlepage}

\section{Introduction}
\label{sec:intro}

Open string instantons are holomorphic maps from Riemann surfaces with
boundaries to the $CY_3$ target space.  It is understood that the
boundaries of the instanton end on special Lagrangian submanifolds of
the threefold. In other words, we are interested in the problem of
counting holomorphic disks with boundary on a Lagrangian submanifold.

In \cite{AV-00}, Aganagic and Vafa used Mirror Symmetry to determine
these open string instantons: the B-model superpotential can be
computed exactly and mapped to the A-model superpotential in the large
volume limit of the $CY_3$.  The latter cannot be computed exactly,
but contains instanton corrections, ie holomorphic disks ending on
``A-model branes'' (also called A-branes); comparison with the B-model
superpotential allow us to determine the contribution of these
instantons and their degeneracy.

The disk amplitude in the large volume limit (ie $e^v \to 0$), which -
in the type II context - has the interpretation of superpotential
corrections to $4d$ $N=1$ susy, is expected to be of the form
\cite{OV-99} (A-model superpotential):
\begin{equation}  \label{A-model}
W_A = \sum_{n \geq 1 \atop k,m} \frac{d_{k,m}}{n^2} q^{nk} y^{nm},  
\end{equation}
where $q=e^{-t}$ and $y=e^v$ are the (exponentiated) closed and open
string complexified K\"ahler classes, measuring respectively the
volume of compact curves and holomorphic disks embedded in the
threefold.  The coefficients $d_{k,m}$ are the numbers of primitive
holomorphic disks labelled by the classes $k$ and $m$ -- two vectors
in the homologies $H_2$ of the threefold and $H_1$ of the brane
respectively.

The tables given by \cite{AV-00} exhibit the integrality of the
coefficients $d_{k,m}$ for their two examples of threefolds: the
resolved confiold and the degenerate $\P \times \P$.  The
current paper proves this for all $k,m$ by analytic means.  We first
Fourier expand the B-model superpotential \cite{AV-00} in $q$ and $y$
and equate it to \ref{A-model}, then we prove the integrality
of $d_{k,m}$ by proceeding inductively on the greatest common divisor
of $k$ and $m$.  We derive interesting congruences of binomial
coefficients modulo powers of a prime.

This explicit method is only possible due to the simple nature of the
mirror map: for $\P$ (both examples), the relation between $t$ and
$\hat t$ is rational: $q=\hat q/ (1+\hat q)^2 $.

The first four sections are a reminder of the method used by
\cite{AV-00}; it rests on the equivalence of the A- and B-model under
mirror symmetry.  On one hand, the A-model string amplitude is
re-interpreted in topological string theory as counting holomorphic
maps from Riemann surfaces (with boundary) to the target space; on the
other hand, the B-model amplitude is obtained via Chern-Simons
reduction to the world-volume of the B-brane.

The last two sections and the mathematical appendix are the crux of
the paper: we prove the integrality of the open instanton numbers in
the examples of the resolved conifold $\CO(-1) \times \CO(-1) \to \P$
and of the degenerate  $\P \times \P$ (where one  $\P$ gets
infinite volume).   The appendix proves the following powerful
congruences for a prime $p>3$ and integers $n,k,l$:
$$
{np^l \choose kp^l} \equiv {n p^{l-1} \choose kp^{l-1}}
~{\rm mod}~ p^{3l}, \qquad
\qquad (p^k-1)!' \equiv -1 ~{\rm mod}~ p^l \quad (k\geq l)
$$
These are generalisations of Wolstenholme's and Wilson's theorems
respectively. 

\paragraph{Acknowledgements:} Thanks to Emanuel Scheidegger for his
unremitting help, to Matthijs Coster for his contribution towards
lemma \ref{bin-cong}, to Pieter Moree for stimulating discussion and
to Robbert Dijkgraaf for the final touch.


\section{The A-model}
\label{sec:The-A-model}

For the A-model, we consider a $U(1)$ linear sigma model, ie a
complex K\"ahler manifold $Y$ obtained by quotienting the
hypersurface
\begin{equation}
  \label{L_Y}
  L_Y:= \big\{ \sum_{i=1}^n Q_i |\Phi^i|^2 = r^2 \big\}
\end{equation}
of $\C^n$ by a $U(1)$ subgroup of the isometry group of $\C^n$.  The
charges $Q_i$ are integers, and if they sum up to $0$, $Y$ is a
complex $(n-1)$ dimensional CY manifold (non-compact, as the
directions with negative charge are non-compact).

More generally, we view   $\C^n$ as a torus fibration $T^n \to L$,
where the base is just $\R^n$ parametrised by the $|\Phi^i|$.  We
can also consider a real $k$-dimensional subset $L_Y$ of $L$ given by
$(n-k)$ equations (\ref{L_Y}) for $(n-k)$ sets of charges  $Q_i^a$,
and then divide the fibration by $U(1)^{n-k}$ to obtain a fibration
$Y= (T^k \to L_Y)$ which is a complex $k$-dimensional non-compact CY
manifold. 

Note that the base $L_Y$ is a Lagrangian submanifold of $Y$: since $L$
is given by fixing values of the arguments $\theta^i$ of the
complex variables $\Phi^i$, the K\"ahler form $\omega =
\sum_{i=1}^n d |\Phi^i|^2 \wedge d \theta^i$  vanishes on it.
The base $L_Y$ is our first example of a D$k$-brane.  

Other examples of D$k$-branes are obtained by considering rational
linear subspaces of $L_Y$, ie submanifolds $D^r \subset L_Y$ of
real dimension $r\leq k$, given by $(k-r)$ constraints
\begin{equation}
  \label{A-brane}
  \sum_{i=1}^n q_i^\alpha |\Phi^i|^2 = c^\alpha 
\end{equation}
with integers $q_i^\alpha$, $\alpha=1,\dots,k-r$.  
Since the slope of $D^r$ is rational, the $(k-r)$-dimensional subspace
of the fibre $T^k$ above any point of $D^r$ and orhtogonal - wrt $\omega$
- to $T_p D^r$ is itself a torus $T^{k-r}$.  That is, we have a new
D$k$-brane given by the fibration $T^{k-r} \to D^r$.  As a
submanifold of $Y$, it is \emph{special} Lagrangian iff $\sum_i
q_i^\alpha = 0 $.  All these special Lagrangian submanifolds of $Y$
are called A-branes.

\section{The B-model}
\label{sec:The-B-model}

The mirror equation to (\ref{L_Y}) is 
\begin{equation}
  \label{B-model}
  \prod_{i=1}^n y_i^{Q_i} = e^{-t}
\end{equation}
where $t:= r+i \theta$ is the complexified K\"ahler parameter, ie the
Fayet-Iliopoulos term $r$ of (\ref{L_Y}) combined with the $U(1)$
$\theta$ angle.  The $y_i$ are homogeneous coordinates for ${\bf
  P}^{n-1}$.  When $L_Y$ is given by a set of $(n-k)$ equations,
(\ref{B-model}) also consists of $(n-k)$ equations for different
K\"ahler parameters $t_a$.

Note that (\ref{B-model}) is not yet the equation of the mirror CY
space.  The B-model is a Landau-Ginsburg theory with superpotential
$$
W(y_i) =\sum_{i=1}^n y_i,
$$
in which $(n-k)$ of the complex variables $y_i$ can be substituted by
(\ref{B-model}), leaving just $k$ of them. The mirror CY space is
compact or not according to whether we add or not a gauge-invariant
superpotential term $PG(\phi_i)$ to the original theory.
In the first case, the CY is given by an orbifold of the hypersurface
$W(y_i)=0$, thus $(k-2)$-dimensional.  In the second case, it is given
by $W(y_i)= xz$, where $x,z$ are affine (and not projective!)
coordinates giving rise to non-compact directions, thus
$k$-dimensional.  (Note that sometimes the $y_i$ variables occuring in
$W(y_i)$ are rescaled to new variables $\tilde{y}_i$ such that these
appear with powers different from $1$.)

As for the B-brane, the mirror of equation (\ref{A-brane}) is 
\begin{equation}
  \label{B-brane}
  \prod_{i=1}^n y_i^{q_i^\alpha} = \epsilon^\alpha
  e^{-c^\alpha}, \qquad \alpha = 1,\dots, k-r,
\end{equation}
as a subspace of the mirror CY.  We have allowed a phase
$\epsilon^\alpha$ to occur; in other words, we have complexified
$c^\alpha$.   Thus the B-brane is a holomorphic submanifold of
complex dimension $k-(k-r)=r$, ie it is a D$(2r)$-brane, where $r$ was
the real dimension of the base of the A-brane.

\section{Topological Strings and Chern-Simons action}

In order to extract instanton numbers from our description of A-branes
and their mirror B-branes, we need an alternative way of computing the
B-model superpotential.  We find salvation in topological string
theory, where the A-model string amplitude counts holomorphic maps
from Riemann surfaces with boundary to the target space with
the boundary ending on A-branes, while the B-model amplitude computes
the holomorphic Chern-Simons action reduced to the world-volume of the
B-brane.  Hence it only works for the CY \emph{three}folds, ie from now on we
restrict to $k=3$.  

Since the A-model disk amplitude in the large volume limit computes
corrections to the $4d$ $N=1$ superpotential, we can extract its instanton
numbers from the B-model superpotential, ie from the classical action 
\begin{equation}
  \label{CS-action}
  W  = \int_Y \Omega \wedge \textrm{ Tr } [A\bar{\partial} A +
  \frac{2}{3} A^3] 
\end{equation}
for a holomorphic $U(N)$ gauge field $A \in H^{0,1} (Y, \textrm{ adj }
U(N))$. 

We shall be interested in the cases where the B-brane is a D2-brane, ie
a holomorphic curve $\CC$; that is the case $r=1$ with $r$ being the
real dimension of the base of the A-brane.  Then the components of the
gauge field $A$ are holomorphic sections of the normal bundle
$N(\CC)$, call them $s$, and the reduced Chern-Simons action is 
\begin{equation}
    W(\CC) = \int_{\CC} \Omega_{ijz} s^i
    \bar{\partial}_{\bar{z}} s^j dz d\bar{z} ,
\end{equation}
which vanishes in the light of $\bar{\partial}_{\bar{z}} s^j (z) =
0$.  This is clearly unattractive for our purposes.  A way of
obtaining a non-vanishing result is to consider the variation of the
integral under holomorphic deformations of $\CC$. This won't vanish
if we have obstructions to holomorphic deformations, such as boundary
conditions for the B-brane at infinity.

This requires a non-compact B-brane $\CC$, hence a non-compact mirror
CY given by $\{W(y_i) = xz\}$ for $k=3$ homogeneous coordinates $y_i$.
This equation reads $\{F(u,v)=xz \}$ for two affine complex variables
$u,v$, say $y_1=e^u, y_2=e^v$.  Since $r=1$, the B-brane is given by
$k-1=2$ equations in the variables $y_i$, hence fixing $W(y_i)$ or
$F(u,v)$ to a constant value.  If this value is $0$, the B-brane will
split into two submanifolds $\{x=0\}$ and $\{z=0\}$ and hence
deformations will be obstructed (as otherwise the brane would pick up
a boundary) and the B-model superpotential $W(\CC)$ will not vanish,
as desired.

Note that we can similarly obtain configurations where the A-brane
will split in two: for instance, a charge $q=(1,-1,0,\dots,0)$
restricts the Lagrangian submanifold $L_Y$ to $\{ |\Phi_1|^2
  -|\Phi_2|^2 =c \}$  and for vanishing $c$ the A-brane will enter a
phase where it splits into $\{ \Phi_1 =\Phi_2 \}$ and $\{ \Phi_1
  =-\Phi_2 \}$.  

To finish off the computation of the B-model Chern-Simons action, we
fix the values of one of the affine parameters $u,v$ of the B-brane at
infinity to some constant value (say $v\to v_*$ for large $|z|$).
This parameter $v$ measures, on the A-model side, the size of the
holomorphic disk ending on the brane.  We then choose $u$ and $v$ as
the two sections of the normal bundle $N(\CC)$.  $\CC$ itself is
parametrised by $z$, and the last variable $x$ parametrising the
B-brane is set to $0$.  We write the holomorphic 3-form as
$\Omega=du dv \frac{dz}{z}$ and obtain for the above integral:
$$
W(\CC) = \int_\CC \frac{dz}{z} u \bar{\partial}_{\bar{z}} v
d\bar{z} =  \int_{v_*}^v u dv.
$$
This has the form of an Abel-Jacobi map for the 1-form $u dv$ on the
Riemann surface $F(u,v)=0$, each point of which parametrises a
different B-brane $\CC$. 

Thus, comparing the A- and B-models:
$$
\partial_v W_B = u = \dots \{ F(u,v)=0 \} \dots \stackrel{!}{=} \partial_v
\Big(\sum_{n\geq1} \sum_{k,m} \frac{d_{k,m}}{n^2} (e^{-t})^{nk}
(e^v)^{nm} \Big) = \partial_v W_A 
$$
and the dots mean that we solve $F(u,v)=0 $ for $u$ to obtain an
expression dependent on $v$ and -- through (\ref{B-model}) -- on
$e^{-t}$.

\section{Appreciation of the AV method}

The method of \cite{AV-00} is quite powerful, as it only requires
knowledge of the mirror CY (specifically of the mirror superpotential
$W(y_i)$ or $F(u,v)$) to extract A-model instanton numbers.  Indeed,
the result of the B-model Chern-Simons action ($\partial_v W_B = u$)
is independent of the mirror CY or the mirror B-brane.  

The drawback is that it is not clear how the instanton numbers depend
on the choice of A-branes.  Would different charges $q^\alpha$ yield
different instantons numbers ?  In their examples, the choices of
$q^\alpha$ yield convenient A- and B-branes.  Maybe a different choice
would not allow for several phases in which the brane splits, or would
not allow us to identify one of the variables $u,v$ with the size of
disk instantons.  It seems that given a mirror CY (or even the A-model
CY for that matter), there is a unique choice of A-brane for which we
can compute A-model instantons.

Another constraint of the method is that it only works for $k=3$, as
it relies on the Chern-Simons theory for the B-model topological
string, which presupposes threefolds as target spaces.

\section{Example: the resolved conifold}

We now turn to a non-compact example of $CY_3$, namely the resolved
conifold: $\CO(-1) \times \CO(-1) \to \P$, a rank two concave bundle
over the complex line.  $H_2$ of the $CY_3$ is thus $H_2(\P)=\Z$,
while the A-brane is a Lagrangian submanifold cutting the base $\P$
in a circle $S^1$, and $H_1(S^1)=\Z$.  Thus both $k$ and $m$ are
merely integers and $t,v$ merely complex numbers.  The input from the
B-model is an explicit expression for the derivative of the
superpotential \cite{AV-00}:
\begin{equation} \label{B-model-eg1}
  \begin{split}
\p_v W &= \log \left( \frac{1-e^v}{2} +\frac{1}{2} \sqrt{(1-e^v)^2 +
  4e^{-t+v}} \right),\\
    &= \log \left( \frac{1-y}{2} +\frac{1}{2} \sqrt{(1-y)^2 +4qy} \right)\\
    &= \sum_{k\geq 0, m\geq k} \frac{(-1)^{k+1}}{m+k} {m+k \choose
  k} {m \choose k} q^k y^m \\
   &=: \sum_{k\geq 0, m\geq k} C_{k,m} q^k y^m \qquad \textrm{with
  } C_{0,0}=0,
  \end{split}
\end{equation}
where $v$ is the (rescaled) natural variable in the phase where the
mirror B-brane degenerates to two submanifolds passing through the
South pole of the resolved conifold.  This $v$ also measures the size
of the minimal holomorphic disk passing through the South pole and
ending on the Lagrangian submanifold.  Precisely when the submanifold
splits into several components can we wrap the A-brane around any of
those, and guarantee that it will not deform (as it would otherwise
acquire a boundary).  This phase is characterised by $e^v \to 0$,
agreeing with the large volume limit on the A-model side.

To detail how we arrived at the Taylor expansion of (\ref{B-model-eg1})
in the large volume limit $e^v \to 0$, it is best to differentiate
both sides wrt $q$ and set $a:= 1+ \frac{4qy}{(1-y)^2}$:
\begin{equation}
  \begin{split}
\frac{2y}{(1-y)^2}\frac{1}{a+\sqrt{a}} &= \frac{2y}{(1-y)^2}
(\frac{1}{\sqrt{a}}-1) \frac{1}{1-a} = \frac{-1}{2q}(\frac{1}{\sqrt{a}}-1) \\
    &= \frac{-1}{2q} \sum_{k\geq 1}{-1/2\choose k}\left(
    \frac{4qy}{(1-y)^2} \right)^k \\
    &= \frac{-1}{2q} \sum_{k\geq 1} q^k (-1)^k 2
    {2k-1\choose k} \sum_{i \geq 0} {2k+i-1\choose i} y^{i+k}\\
    &= \sum_{k\geq 0, m\geq k} q^k y^m (-1)^k {2k+1\choose
      k}{m+k \choose 2k+1} \\
    &= \sum_{k\geq 0, m\geq k} q^k y^m (-1)^k {m\choose
      k+1}{m+k \choose k} 
  \end{split}
\end{equation}
And this agrees with the above: 
\begin{equation} \label{Ckm}
  C_{k,m}= \frac{(-1)^{k+1}}{k} {m+k-1 \choose k-1} {m \choose k} 
       = \frac{(-1)^{k+1}}{m+k} {m+k \choose k} {m \choose k}
\end{equation}
As far as the constant of integration is concerned (the $q^0$ term
of (\ref{B-model-eg1})), note that $ \sum_{m\geq 0} C_{0,m} y^m =
\sum_{m\geq 1} \frac{-y^m}{m} = \log (1-y)$, in agreement with the
first expression of (\ref{B-model-eg1}) which goes like $\log (1-y+O(q)) =
\log (1-y) +O(q)$.

Comparing this to the A-model expression (\ref{A-model})
$$
\p_v W = - \sum_{k,m} m d_{k,m} \log (1-q^k y^m) 
        = \sum_{k,m} \left( \sum_{l|(k,m)}
          d_{\frac{k}{l},\frac{m}{l}} \frac{m}{l^2} \right) q^k y^m ,
$$
we can recursively extract the values of all $d_{k,m}$ from the
relation
\begin{equation}  \label{Ckm-dkm}
 C_{k,m} = \sum_{l|(k,m)} d_{\frac{k}{l},\frac{m}{l}} \frac{m}{l^2}.
\end{equation}

\begin{prop}
  With the $C_{k,m}$ as in (\ref{Ckm}), the instanton numbers
  $d_{k,m}$ are all integers.
\end{prop}
\begin{proof}
We proceed step by step, according to the greatest common divisor (gcd)
of $k$ and $m$:

\underline{$(k,m)=1$:} From (\ref{Ckm-dkm}) we have  $C_{k,m} = d_{k,m}
m$.  So for $d_{k,m}$ to be integer, we need $C_{k,m}$ to be $0$ mod
$m$.  Note that in general, $(n,k)=1$ implies $n | {n \choose k}$,
since ${n \choose k} =  \frac{n}{k} {n-1 \choose k-1}$.  Thus
$C_{k,m} \in \Z$ and even $\in m \Z$.

\underline{$(k,m)=p^l$:} for $p$ prime.  This time 
\begin{equation}
  \begin{split}
C_{k,m} =& d_{k,m} m + d_{\frac{k}{p},\frac{m}{p}} \frac{m}{p^2} +
 \dots + d_{\frac{k}{p^l},\frac{m}{p^l}} \frac{m}{p^{2l}} \\
       =& d_{k,m} m + \frac{1}{p} C_{\frac{k}{p},\frac{m}{p}}
  \end{split} 
\end{equation}
Thus for $d_{k,m}$ to be integer, we need $ C_{k,m} \equiv
\frac{1}{p} C_{\frac{k}{p},\frac{m}{p}}$ mod $m$, ie. $ p C_{p^l k, p^l
  m} \equiv C_{p^{l-1} k, p^{l-1} m} $ mod $m p^{l+1}$ for
$(k,m)=1$, ie. 
$$
{p^l (m+k) \choose p^l k} {p^l m \choose p^l k} 
- {p^{l-1} (m+k) \choose p^{l-1} k} {p^{l-1} m \choose p^{l-1} k}
\equiv 0 \textrm{ mod } m p^{2l}.
$$  
Lemma \ref{bin-cong} tells us that the congruence is
valid mod $p^{3l}$ ($p>3$) or mod $p^{3l-1}$ ($\forall p$), hence
also mod $p^{2l}$ for any prime $p$.   Since $m | {p^l m \choose
  p^l k}$, both terms also contain a factor of $m$, and the congruence
is valid mod $mp^{2l}$.

\underline{$(k,m)=pq$:} for primes $p$ and $q$.  Again by (\ref{Ckm-dkm})
we have
\begin{equation}\label{pq}
  \begin{split}
C_{k,m} =& d_{k,m} m + d_{\frac{k}{p},\frac{m}{p}} \frac{m}{p^2} 
         + d_{\frac{k}{q},\frac{m}{q}} \frac{m}{q^2} 
         + d_{\frac{k}{pq},\frac{m}{pq}} \frac{m}{p^2q^2} \\
   =& d_{k,m} m + \frac{1}{p} C_{\frac{k}{p},\frac{m}{p}} +
         \frac{1}{q} C_{\frac{k}{q},\frac{m}{q}} 
         - \frac{1}{pq} C_{\frac{k}{pq},\frac{m}{pq}} 
  \end{split} 
\end{equation}
Thus we need $ pq C_{pqk,pqm} - q C_{qk,qm} -p C_{pk,pm} + C_{k,m}
\equiv 0$ mod $m p^2 q^2$ for $(k,m)=1$, ie 
$$
{pq (m+k) \choose pq k} {pq m \choose pq k} 
- {q(m+k) \choose qk} {qm \choose qk} - {p(m+k) \choose pk} {pm \choose pk} 
+ {m+k \choose k} {m \choose k} \equiv 0 \textrm{ mod } m q^2 p^2. 
$$  
Again by Lemma \ref{bin-cong}, the first difference is $0$ mod $p^3$
($p>3$, or mod $p^2$ $\forall p$),
so is the last difference, and we can factor out $p^2$, hence also $q^2$.
As before, we can also take out a factor of $m$, and the whole line is
thus $0$ mod $m p^2 q^2$.

\underline{$(k,m)=pqr$:} for primes $p,q$ and $r$.  As before, the
principle of inclusion and exclusion yields the requirement
$$
pqr C_{pqrk,pqrm} - qr C_{qrk,qrm} - pr C_{prk,prm} - pq
C_{pqk,pqm} + r C_{rk,rm} + q C_{qk,qm} + p C_{pk,pm} - C_{k,m} 
\equiv 0 \textrm{ mod } m p^2 q^2 r^2
$$
for $(m,k)=1$.  Reasoning as above and noting that the four pairs $ (pqr
C_{pqrk,pqrm} - qr C_{qrk,qrm}) $, $( pr C_{prk,prm} - r C_{rk,rm})$, $(pq
C_{pqk,pqm} - q C_{qk,qm})$ and $(p C_{pk,pm}-C_{k,m})$ are all $0$ mod
$p^2$, we find that the requirement is met.

\underline{$(k,m)=p^l q$:} for $p,q$ prime.  Now we have 
$$
C_{k,m} = d_{k,m} m + \frac{1}{p} C_{\frac{k}{p},\frac{m}{p}}
 + \frac{1}{q} C_{\frac{k}{q},\frac{m}{q}} - \frac{1}{pq} 
 C_{\frac{k}{pq},\frac{m}{pq}},
$$
so we are back at a combination of the cases $(k,m)=p^l$ and
$(k,m)=pq$, and the same reasoning will show that $d_{k,m}$ is again
integer. 

Having covered the cases of $(k,m)$ being product of primes and powers of
primes, inductive reasoning will show that the same conclusion will be met
in the most general case where $(k,m)=p_1^{l_1} \dots p_j^{l_j}$.
\end{proof}

\section{Example: Degenerate $\P \times \P$}

Our second example of non-compact $CY_3$ is a concave line bundle over
two complex lines: $\CO (-3) \to \P \times \P$, with K\"ahler moduli $t_1,
t_2$ describing the sizes of the two complex lines (or real spheres).
An easy mirror map is only known for the degenerate case where the
size of the second $\P$ goes to infinity; that is we retain only one
modulus, $t_1$, with associated variable $q=e^{-t_1}$.  And so -- as in
the previous example - $k$ and $m$ are both integers.

This time the input from the B-model is \cite{AV-00}:
\begin{equation} \label{B-model-bis}
  \begin{split}
\p_v W &= \log \left( \frac{1+q-y}{2} + \frac{1}{2} \sqrt{(1+q-y)^2 -4q}
\right) \\
    &= \sum_{k \geq 0, m\geq 1} \frac{-1}{m}{m+k-1 \choose
      k}^2 q^k y^m\\ 
   &=: \sum_{k,m\geq 0} C_{k,m} q^k y^m \qquad \textrm{with
  } C_{k,0}=0,
  \end{split}
\end{equation}
where $v$ is the (rescaled) natural variable in the phase where the
projection of the A-brane on the base is a circle on the $\P$ of
infinite volume.  In order to understand the double series expansion,
we proceed as in the previous example, but now we differentiate both
sides wrt $y$ and obtain -- up to a minus sign -- something symmetric
in $q$ and $y$:
\begin{equation}
  \begin{split}
    \frac{1}{\sqrt{(1+q-y)^2-4q}} &= \frac{1}{\sqrt{(1-q-y)^2-4qy}}\\
    &= \frac{1}{1+q-y} \sum_{n\geq 0} {-1/2\choose n}\left(
    \frac{-4q}{(1+q-y)^2} \right)^n\\
    &= \sum_{n\geq 0} {-1/2\choose n} (-4q)^n \sum_{i\geq
    0} {i+2n\choose i} (y-q)^i\\
    &= \sum_{m\geq 0} y^m \sum_{n\geq 0} {-1/2\choose n}
    (-4q)^n \sum_{i\geq 0} {m+i+2n\choose m+i} {m+i\choose n} (-q)^i\\
    &= \sum_{m\geq 0} y^m \sum_{k\geq 0} q^k (-1)^k
    \sum_{n=0}^k {-1/2\choose n} 4^n {m+k+n\choose m+k-n}
    {m+k-n\choose m}\\
    &= \sum_{m\geq 0} y^m \sum_{k\geq 0} q^k (-1)^k
    \sum_{n=0}^k 2 (-1)^n \frac{(2n-1)!}{(n-1)!n!}
    \frac{(m+k+n)!}{(2n)!m!(k-n)!} \\
    &= \sum_{m\geq 0} y^m \sum_{k\geq 0} q^k (-1)^k
    {m+k\choose k} \sum_{n=0}^k (-1)^n {m+k+n\choose n} {k
    \choose n}\\
    &= \sum_{m\geq 0} y^m \sum_{k\geq 0} q^k
    {m+k\choose k}^2,
  \end{split}
\end{equation}
where we have used: the last sum over $n$ is but the contribution to
the power $x^k$ in the expansion of the product of $(x-1)^k$ and
$(\frac{1}{1-x})^{m+k+1}$; and since this product equals $(-1)^k
(1+x+x^2+\dots)^{m+1}$, the sum equals $(-1)^k {m+k \choose k}$.

And this agrees with the $C_{k,m}$ above\footnote{Note that in \cite{AV-00}, the $C_{k,m}$ have the following
  form: 
$$
C_{k,m} = \frac{(-1)^{k+1}}{m+k} {m+k\choose k} + \sum_{n=1}^k
\frac{(-1)^{k+n+1}}{m+k+n} \ \frac{(m+k+n)!}{n!n!m!(k-n)!} 
$$
where the first term is just the $n=0$ term of the sum next to it and
is the coefficient in the expansion of $\log (1+q-y)$, so
that the expression agrees with our own one.} : 
\begin{equation} \label{Ckm-bis}
  \begin{split}
C_{k,m} &= -\frac{1}{m} {m+k-1 \choose k}^2 \\
        &= -\frac{1}{k} {m+k-1 \choose k} {m+k-1 \choose  k-1} \\
       &= -\frac{m}{(m+k)^2} {m+k \choose m}^2,
  \end{split}
\end{equation}
of which only the last version is suitable for the case $m=0$:
$C_{k,0}=0$.  The latter yields also the constant of integration (the
$y^0$ term of (\ref{B-model-bis})), since
$ \sum_{k\geq 0} C_{k,0} q^k = 0$, in agreement with the first
expression of (\ref{B-model-eg1}) which goes like $\log
(\frac{1+q-y}{2}+ \frac{1-q}{2}+O(y)) = \log (1+O(y)) = O(y)$.

\begin{prop}
  With the $C_{k,m}$ as in (\ref{Ckm-bis}), the instanton numbers
  $d_{k,m}$ are all integers.
\end{prop}
\begin{proof}
  The logic remains the same, and we proceed again inductively on the
  nature of the gcd of $k$ and $m$.

\underline{$(k,m)=1$:} As before, we need $C_{k,m} \equiv 0$ mod $m$,
  which is readily seen from (\ref{Ckm-bis}). 

\underline{$(k,m)=p^l$:} As before, the requirement boils down to
$p C_{p^l k, p^l m} \equiv C_{p^{l-1} k, p^{l-1} m}$ mod $mp^l$ for
$(k,m)=1$, ie
$$
\frac{m}{(m+k)^2} \left[ {p^l (m+k) \choose p^l m}^2  
- {p^{l-1} (m+k) \choose p^{l-1} m}^2 \right] \equiv 0 \textrm{ mod }
m p^{2l},
$$  
ie
$$
{p^l (m+k) \choose p^l m} \equiv \pm {p^{l-1} (m+k) \choose
  p^{l-1} m} \equiv 0 \textrm{ mod } p^{2l},
$$
which is again fine by Lemma \ref{bin-cong}.

\underline{$(k,m)=pq$:} The same requirement as in (\ref{pq}) stipulates
$$
{pq (m+k) \choose pq k}^2 - {q(m+k) \choose qk}^2 - {p(m+k) \choose
  pk}^2 + {m+k \choose k}^2 \equiv 0 \textrm{ mod } q^2 p^2.  
$$ 
for $(k,m)=1$.  Again, by Lemma \ref{bin-cong}, the first
difference is $0$ mod $p^3$, so is the second, and similarly for mod
$q^3$.

The cases \underline{$(k,m)=pqr$} and \underline{$(k,m)=p^l q$} can be
imported without change from the previous example, and thus the integrality
of the $d_{k,m}$ is proved for the most general case of
$(k,m)=p_1^{l_1}\dots p_j^{l_j}$. 
\end{proof}

\newpage
\appendix
\section{Appendix}

We now prove a lemma from number theory, involving congruences of
binomial coefficients: 
\begin{lem} \label{bin-cong} For $p$ prime and $n,k \in \N$ we have
\[
{np^l \choose kp^l} \equiv {n p^{l-1} \choose kp^{l-1}}
~{\rm mod}~ p^{3l} \quad \textrm{ for } p>3, (and ~{\rm mod}~ p^{3l-1} 
\ \forall p) 
\]
\end{lem}

\begin{proof}
We use the notation $\prod', \sum'$ for a product or a sum
skipping multiples of $p$, and we define $S(n):=\sum_{i=1}^{'n} 
\frac{1}{i}$ and $S_2(n):=\sum_{i=1}^{'n} \frac{1}{i^2}$. Note 
also that all non-multiples of $p$ have an inverse, ie that $(\Z
/p^l \Z )^*$ is a multiplicative group.  We have
  \begin{equation}
    \begin{split}
      \prod_{i=1}^{kp^l} { }^{'} (1+\frac{kp^l}{i}) &= \prod\nolimits'
      \frac{kp^l+i}{i} \ \frac{kp^l-i}{i} \\ 
       &= \prod\nolimits' (1-\frac{k^2 p^{2l}}{i^2}) \equiv 1+ k^2
      p^{2l} S_2(kp^l) \textrm{ mod } p^{4l}
    \end{split}
  \end{equation}
except for an extra minus sign for the second line if $p=2$, $l=1$, $k$ odd.
The lhs is $ 1+ S(kp^l) kp^l - \frac{S^2(kp^l)-S_2(kp^l)}{2} k^2 p^{2l}$
mod $p^{3l}$.  Comparing both sides mod $p^{2l}$, we find that
$S(kp^l) \equiv 0$ mod $p^l$.  Comparing mod $p^{3l}$ yields $k
S(kp^l) + \frac{k^2}{2} S_2(kp^l) p^l \equiv 0$ mod $p^{2l}$; and using
$S_2(kp^l) \equiv 0$ mod $p^l \quad (p>3)$ from Lemma \ref{S_n}, we obtain 
$$
S(kp^l) \equiv 0 \textrm{ mod } p^{2l} \textrm{ for } p>3,
$$
while only mod $p^{2l-1}$ for $p=3$, and mod $p^{2l-2}$ for $p=2$ (as the
coefficient $\frac{1}{2}$ takes away one power of $p$).  

We now turn to the binomial coefficients: Note first that they both
have the same number of multiples of $p$, namely the number of
multiples of $p$ lying in the interval $[n-k,n]$ or $[k,n]$ --
whichever interval is smaller.  We assume that they actually do not
contain multiples of $p$, so that we can consider their quotient.  If
they do, their difference will contain even more powers of $p$ than
$p^{3l}$, so that we could strengthen our result.
   \begin{equation}
    \begin{split}
{n p^l \choose kp^l } / {n p^{l-1} \choose kp^{l-1} } &= \frac{n
  p^l\dots ((n-k)p^l +1)}{n p^{l-1} \dots ((n-k)p^{l-1} +1)} \
  \frac{(kp^{l-1})!}{(kp^l)!}  \\
&= \prod_{i=1}^{kp^l} { }^{'} \frac{(n-k)p^l+i}{p^{-kp^{l-1}}} \
  \frac{p^{-kp^{l-1}}}{kp^l-i}  \\
&= \prod\nolimits' \frac{(n-k)p^l+i}{i} 
= \prod\nolimits' (1+\frac{(n-k)p^l}{i})\\ 
&\equiv 1+ p^l (n-k) S(kp^l)+ p^{2l} (n-k)^2 \frac{S^2(kp^l)-S_2(kp^l)}{2}
  \textrm{   mod } p^{3l} \\
&\equiv 1 \textrm{  mod } p^{3l}
    \end{split}
  \end{equation}
by the above.  
\end{proof}

As a special case of the lemma, for $n=2$, $k,l=1$, we obtain ${2p
  \choose p} \equiv 2$ mod $p^3$, or Wolstenholme's theorem: 
\begin{cor}
\quad $ {2p-1 \choose p-1} \equiv 1$ {\rm mod} $ p^3$ for $p>3$,
(and {\rm mod} $p^2 \quad \forall p$).
\end{cor}

\begin{lem} \label{S_n}
  For $l,n\in\N$ and $p$ prime we have
  $$
S_n(p^l):= \sum_{i=1}^{p^l} { }^{'} \frac{1}{i^n} \equiv 0 \equiv
\sum_{i=1}^{p^l} { }^{'} i^n ~{\rm  mod}~ p^l \qquad \textrm{ if }
(p-1)\nmid n, 
$$
and $0$ {\rm mod} $p^{l-1}$ for any $p,n$.
\end{lem}

\begin{proof}
  Note that the same is true of $S_n(kp^l)$ for $k\in\N$, as this is
  merely $k$ copies (mod $p^l$) of $S_n(p^l)$.  Similarly, $S_n(p^{l+1})$
  is just $p$ equal copies (mod $p^l$) of $S_n(p^l)$, so by induction, we
  only need to prove the result for $S_n(p)$.

  Let $\zeta$ be a primitive root mod $p$, ie a number such that the
  set $\{1,\zeta,\zeta^2,\dots,\zeta^{\phi-1} \}$ covers all elements
  of the multiplicative group $(\Z/p\Z)^*$ of order $\phi(p)=p-1$.
  That is, the set is equal (mod $p$) to $\{1,2,\dots,p-1 \}$; and
  similarly the set $\{1,\frac{1}{2^n},\dots,\frac{1}{(p-1)^n} \}'$ is
  equal (mod $p^l$) to $\{1,\zeta^n,\zeta^{2n},\dots,\zeta^{n(p-2)}
  \}$.  Hence
$$
S_n(p) \equiv 1+ \zeta^n+ \dots+\zeta^{n(p-2)} =
  \frac{1-\zeta^{n(p-1)}}{1-\zeta^n} \equiv 0 \textrm{ mod } p
$$ 
since $\zeta^{p-1} \equiv 1$ mod $p$.  For the denominator $1-\zeta^n$
to be invertible mod $p$, we must exclude the case where
$\zeta^n\equiv 1$ mod $p$, ie where $n$ is a multiple of
$p-1$.  In this case, it still is true that $S_n(p)\equiv 0$
mod $p^0$, ie 0 mod $1$.

For $ \sum' i^n$, the proof runs similarly.  Note that in this case we
could drop the dash from the sum to include multiples of $p$, as their
contribution would be $p(1+2+\dots+p^{l-1}) = p^l (p^{l-1}+1)/2$.
\end{proof}

One could have tackled the proof of Lemma \ref{bin-cong} in other
ways, in particular by writing out the binomial coefficients as
factorials and using properties of factorials.  For the sake of
completeness, we include a useful property of residues of factorials
(Wilson's theorem):
\begin{prop}
For $p$ prime we have
 $$
(p-1)! \equiv -1  ~{\rm mod}~ p \qquad (p>2),
$$
and $1$ mod $p$ for $p=2$.  
\end{prop}
\begin{proof}
  In the product $1\dots (p-1)$, the numbers occur in pairs $j$ and
  $1/j$ mod $p$, except for $1$ and $p-1$ which are their own
  inverses, since these are the only solutions of $j^2 -1 \equiv 0$
  mod $p$.  Thus the product is $1 (p-1) \equiv -1$ mod $p$.  For
  $p=2$, $1$ and $p-1$ are equal mod $p$.
\end{proof}
For higher powers of the prime $p$, $p^k!$ contains a factor of
$p^{1+p+p^2+\dots+p^{k-1}}$.  We introduce the dash notation to
indicate that we have skipped all these multiples of $p$: $p^k !'= p^k
! /(p^{k-1}!(p)^{p^{k-1}})$.  We compute the residue mod $p$:
  $(p^k-1)!'= (1\dots p^{k-1} \dots 2p^{k-1} \dots pp^{k-1})'$
  consists of $p$ times $(p^{k-1}-1)!'$ mod $p$.  By induction, this
  yields:
\begin{lem}
  For $p$ prime and $k\in \N$ we have
$$
(p^k-1)!' \equiv -1 ~{\rm mod}~ p \qquad (p>2),
$$ 
and $1$ mod $p$ for $p=2$.
\end{lem}

More generally, this result holds also mod $p^l$ for powers $k\geq
l$, as we shall show below.

\begin{lem}
 For $p$ prime we have:
$$
(p^{k-1} -1)!' \equiv (p-1)!^{p^{k-2}} \equiv -1+n_1 p^{k-1} ~{\rm
  mod}~ p^k \qquad (p>2, k\geq 2),
$$
and $\equiv 1 + p^{kl-1}$ mod $p^k$ for $p=2$, $k\geq 4$.

Here, $n_1 \in \Z_p $ is defined by $(p-1)! \equiv -1+n_1 p$ mod $p^2
~(p>2)$.
\end{lem}
\begin{proof}
  By induction on $k$. The case $k=2$ is trivial.
$$
(p^{k-1}-1)!'= \big[1\cdot 2\dots (p^{k-2}-1)\big]' ~\big[(p^{k-2} +1)\dots
(p^{k-2}+p^{k-2}-1)\big]' \dots \big[((p-1)p^{k-2} +1)\dots (p^{k-1}-1)\big]'
$$ 
The first square bracket is $-1+n_1 p^{k-2}$ mod $p^{k-1}$ by
induction; ie it is $-1+n_1 p^{k-2} + c p^{k-1}$ mod $p^k$ (for some
integer $c$), a quantity we denote by $a$.  The second square bracket
is $a + p^{k-2} (p^{k-2}-1)!' S_1(p^{k-2})$ mod $p^k$. Since
$S_1(p^{k-2}) \equiv 0$ mod $p^{k-2}$ by lemma \ref{S_n} ($p\neq
2$), this is just $a$ mod $p^k$ if $k>3$. (For $k=3$, a trailing
$p^2\cdot$const won't affect the ultimate conclusion). All remaining
brackets are also $a$ mod  $p^k$. Hence
$$
(p^{k-1}-1)!' \equiv a^p \equiv (-1 +n_1 p^{k-2})^p \equiv -1
+n_1 p^{k-1} \equiv (-1+n_1 p)^{ p^{k-2}} ~{\rm mod}~p^k.
$$

For $p=2$, the anchor is at $k=4$: $(p^3-1)!'=1~3~5~7 \equiv 1+2^3$
mod $p^4$.  So the last line reads $a^p \equiv 1+p^{k-1}$ mod $p^k$.
Since we only have $S_1(p^{k-2}) \equiv 0$ mod $p^{k-3}$, there is a
trailing $p^{2k-5}$, which is fine for the induction with $k\geq 5$.
\end{proof}

\begin{cor}
  $$ (p^k-1)!' \equiv -1 ~{\rm mod}~ p^k \qquad (p>2) $$
and $1$ mod $p^k$ for $p=2$.
\end{cor}
\begin{proof}
  lhs $= \big[1\dots (p^{k-1}-1)\big]' ~\big[(p^{k-1} +1)\dots
  (p^{k-1}+p^{k-1}-1)\big]' \dots \big[((p-1)p^{k-1} +1)\dots (p^k-1)\big]'$.
  By the previous lemma, the first square bracket yields
  $(p-1)!^{p^{k-2}}$  ($p>2$), while the second yields the same plus
  $p^{k-1} (p-1)! S_1(p)$ (which is 0 mod $p^k$), and all other square
  brackets yield the same. In all we have $(p-1)!^{p^{k-1}}\equiv
  (-1+n_1 p+\dots)^{p^{k-1}} \equiv -1$ mod  $p^k$ (or +1 for $p=2$). 
\end{proof}
The same method of proof easily yields:
\begin{prop}
For prime $p$ and integers $k\geq l$ we have
$$ (p^k-1)!' \equiv -1 ~{\rm mod}~ p^l \qquad  (p>2) $$
and $1$ mod $p^l$ for $p=2$.
\end{prop}

There is no explicit formula for $(p-1)!$ mod $p^2$, ie the integer
$n_1$ in $(p-1)! \equiv -1 + n_1 p$ mod $p^2$ is no evident function
of $p$.  In Hardy and Wright, one will find a formula reducing the
factorial to terms involving $\frac{p-1}{2}!$.  Also, for mod $p^3$,
the recent literature exhibits ways to reduce the factorial to
complicated terms involving the class number of $p$.


\begin{thebibliography}{999}

\bibitem[AV-00]{AV-00} M. Aganagic and C. Vafa, {\it Mirror Symmetry,
    D-Branes and Counting Holomorphic Discs}, hep-th/0012041

\bibitem[OV-99]{OV-99} H. Ooguri, C. Vafa {\it Knot Invariants and Topological
  Strings}, Nucl.Phys. {\bf B577} (2000) 419-438,  
   hep-th/9912123

\end{thebibliography}
\end{document}